\documentclass{cup-hpl}

\usepackage{graphicx}
\usepackage{dcolumn}
\usepackage{bm}
\usepackage{upgreek}
\usepackage[utf8]{inputenc}
\usepackage[T1]{fontenc}
\usepackage{etoolbox}
\usepackage{makecell}
\usepackage{booktabs}

\begin{document}

\newtheorem{theorem}{Theorem}

\shorttitle{An optimized online filter stack spectrometer}                                   
\shortauthor{J.-x. Wen et.al.}

\title{An optimized online filter stack spectrometer}

\author[1,2]{Jia-xing Wen}
\author[3,2]{Ge Ma}
\author[1]{Ming-hai Yu}
\author[1]{Yu-chi Wu}
\author[1]{Yong-hong Yan}
\author[1]{Shao-yi Wang}
\author[3,2]{Huai-zhong Gao}
\author[4]{Lu-shan Wang}
\author[4]{Yu-gang Zhou}
\author[4]{Qiang Li}
\author[1]{Yue Yang}
\author[1]{Fang Tan}
\author[1]{Xiao-hui Zhang}
\author[1]{Jie Zhang}
\author[1,2]{Wen-bo Mo}
\author[1]{Jing-qin Su}
\author[1]{Wei-min Zhou}
\author[1]{Yu-qiu Gu}
\author[1]{Zong-qing Zhao\corresp{Z.-q. Zhao. Laser Fusion Research Center, CAEP, Mianyang, Sichuan, 621900, China. \email{zhaozongqing99@caep.ac.cn}}}
\author[3,2]{Ming Zeng}

\address[1]{Science and Technology on Plasma Physics Laboratory, Laser Fusion Research Center, CAEP, Mianyang, 621900, Sichuan, China}
\address[2]{Department of Engineering Physics, Tsinghua University, Beijing, 100084, China}
\address[3]{Key Laboratory of Particle and Radiation Imaging (Tsinghua University), Ministry of Education, Beijing, 100084, China}
\address[4]{School of Information Engineering, Southwest University of Science and Technology, Mianyang, 621010, China}

\begin{abstract}
The spectrum of laser-plasma-generated X-rays is very important as it can characterize electron dynamics and also be useful for applications, and nowadays with the forthcoming high-repetition-rate laser-plasma experiments, there is a raising demand for online diagnosis for the X-ray spectrum. In this paper, scintillators and silicon PIN diodes are used to build a wideband online filter stack spectrometer. The genetic algorithm is used to optimize the arrangements of the X-ray sensors and filters by minimizing the condition number of the response matrix, thus the unfolding error can be significantly decreased according to the numerical experiments. The detector responses are quantitatively calibrated by irradiating the scintillator and PIN diode using different nuclides and comparing the measured $\gamma$-ray peaks. Finally, a 15-channel spectrometer prototype has been implemented. The X-ray detector, front-end electronics, and back-end electronics are integrated into the prototype, and the prototype can determine the spectrum with 1 kHz repetition rates.
\end{abstract}


\maketitle

\section{INTRODUCTION}
\label{Sec01}
The laser-driven plasma-based electron accelerators and X-ray sources powered by ultra-intense laser technology have obtained extensive research in recent years. With the accelerating gradient of $\sim$100 GeV/m in the laser-plasma accelerators, electrons can be accelerated to hundreds of MeV in a few millimeters\cite{esarey_physics_2009}. Then, these electrons can generate X-rays through betatron radiation\cite{esarey_synchrotron_2002}, inverse Compton scattering\cite{ta_phuoc_all-optical_2012,sarri_ultrahigh_2014,yan_high-order_2017}, and bremsstrahlung\cite{cipiccia_tuneable_2012}, thus providing tabletop complements to large-scale conventional accelerator-based X-ray sources. These X-rays sources have advantages of femtosecond duration, micron source size, wide spectral range\cite{corde_femtosecond_2013}, thus have tremendous potentials for applications\cite{albert_applications_2016}, e.g., biology radiagraphy\cite{guo_high-resolution_2019}, non-destructive testing\cite{jones_evaluating_2016,yang_design_2019}, and high-energy-density physics\cite{tommasini_short_2017,tian_radiography_2019}.

The research and application of the laser-driven tabletop X-ray sources require a unique set of diagnostics\cite{downer_diagnostics_2018}, among which the X-ray spectrometer is a very important one since the X-ray spectrum characterizes the electron dynamics in plasma\cite{kneip_observation_2008,kneip_bright_2010,sarri_ultrahigh_2014,yan_high-order_2017} and is useful for the applications\cite{tian_radiography_2019}. There are multiple types of spectrometers have been developed to cover the photon energies ranging from a few keV to tens of MeV in laser-plasma experiments. For X-rays below 30 keV, the scientific charge-coupled devices (CCD) operated in single-photon counting mode can provide a high-resolution X-ray spectrum by identifying individual X-ray photon signals in one frame of image\cite{yan_calibration_2013,stoeckl_operation_2004}. For X-rays below 100 keV, the crystal spectrometer can be applied, and the spectrum can be measured by detecting the angular distribution of X-rays diffracted by the crystal\cite{yu_hard_2016,chi_diffraction_2017}. For X-rays with energies of hundreds of keV, the electron-track-based spectrometer has been proposed. The X-ray energies can be reconstructed by the detection of Compton electron tracks and energies using silicon trackers\cite{wen_diagnostics_2021}. For X-rays above 1 MeV, the Compton spectrometer can convert the X-rays to electrons by forward Compton scattering. The spectrum of Compton electrons can be measured by a magnetic spectrometer, and the spectrum of X-rays can then be derived\cite{singh_compact_2018,cipiccia_compton_2013,gehring_determining_2014,corvan_design_2014,haden_high_2020}. The spectrometers mentioned above are sensitive to X-rays in relatively narrow energy ranges, instead, the filter stack spectrometer (FSS) can diagnose the X-ray spectrum in wideband, typically from tens of keV to hundreds of MeV. FSS usually uses a stack of filters interlaced with X-ray sensors\cite{courtois_characterisation_2013,jeon_broadband_2015,hannasch_compact_2021}, and the characteristic parameters of the X-ray spectra can be reconstructed from the response matrix (RM) and sensor signals. Simultaneously, FSS has the advantages of simple manufacturing and easy operation, thus has been widely used in laser-plasma experiments.

Many current FSSs use image plate (IP) as the X-ray sensors\cite{courtois_characterisation_2013,jeon_broadband_2015,hannasch_compact_2021}, as IP is reusable, highly sensitive to X-rays, versatile, and resistant to electromagnetic pulse\cite{meadowcroft_evaluation_2008,bonnet_response_2013}. But the IP signals must be read out offline by an IP scanner with a processing time of ten minutes, which is not in line with the high-repetition-rate laser-plasma experiments. Since the high repetition rate is important for the research of the experimental laws, the optimization and application of the X-ray sources\cite{hatfield_data-driven_2021, ma_accelerating_2021}, there is a requirement for online diagnosis for the wideband spectrum of laser-induced X-rays.

The scintillators and semiconductors are commonly used X-ray sensors\cite{grupen_particle_2008}. Scintillators can transfer X-rays to visible scintillation light which can be then transferred into an electric signal easily using a photoelectric converter. Semiconductors can transfer X-rays to charge carriers (electrons and holes). While the charge carriers drift under a bias voltage, electric signals would be induced on the electrode. Since the decay time of scintillators and charge carrier drift time of semiconductors are typically from nanoseconds to hundreds of nanoseconds, and the electric signals can be processed online using an electric readout system, the FSS built by scintillators or semiconductors would have a fast repetition rate in line with the current state-of-the-art and future laser systems. Recently, some online FSSs based on scintillators have been developed\cite{rusby_novel_2018,behm_spectrometer_2018,stransky_development_2021,istokskaia_experimental_2021}, where the charge-coupled devices (CCDs) or complementary metal-oxide-semiconductor (CMOS) cameras are used as the photoelectric converters. We note that the scintillator-based FSS has already been used in the leading-edge experiments\cite{cole_experimental_2018,poder_experimental_2018,underwood_development_2020}, showing excellent performance in the detection of wideband X/$\gamma$-rays with high repetition rate. However, there are some aspects of the currently developed online FSS technique that need improvement. Firstly, the light collection efficiency is low since the lens of CCD or CMOS only covers a small solid angle, which could lead to high statistical uncertainties, especially in the low X-ray fluence condition. Secondly, the scintillation light collection efficiency could be hardly calibrated and easily changed, because the light couplings are not solid. Thirdly, the FSS configuration has been hardly optimized leading to a severely ill-conditioned RM. These existing problems might lead to a relatively large error in the determination of the X-ray spectrum.

In this paper, we propose and implement an optimized online FSS. To alleviate the ill condition of RM, the genetic algorithm\cite{goldberg_genetic_1989} is used to optimize the arrangements of the sensors and filters. To achieve a solid light coupling and a high light collection efficiency, PIN diodes with a large sensitive area and low noise are used as photoelectric converters. PIN diodes are also used as the sensors in the front few channels (close to the X-ray sources) to detect the low-energy X-rays directly, to improve the unfolding accuracy in the tens of keV to one hundred keV range. This configuration promises a direct calibration of the light collection efficiency as well as an alleviated ill condition of the RM, and thus the unfolding error can be decreased to approximately 16\% for a typical spectrum according to the numerical experiments. The electric signals from PIN diodes are amplified by the homemade trans-impedance amplifiers (TIAs) and main amplifiers and then digitalized by a multi-channel data acquisition (DAQ) system based on DRS4 chip\cite{bitossi_ultra-fast_2016}. The filters and X-ray sensors, front-end electronics (FEE), and DAQ are integrated into the online FSS, which is compact and easy to operate. And the online FSS can provide the capability of unfolding the X-ray spectrum from tens of keV to hundreds of MeV with 1 kHz repetition rates.

\section{METHODOLOGY}
\label{Sec02}
\subsection{Spectrometer optimization}
\label{Sec02_SubSec01}
The ill-posed linear system is the most essential drawback FSS suffers from, leading to significant spectrum measurement errors. Since the scintillator could be treated as both the filter and sensor, the layer thickness, number of layers, density of filter, and density of scintillator are expected to be finely tuned to alleviate the ill condition of the RM. Rusby optimized their online FSS according to the difference of the scintillator output between the energy ranges of interest\cite{rusby_novel_2018}. The optimum scintillator should have the largest difference in the scintillator output. But this optimization criteria was spectrum-dependent and the thicknesses of scintillators were kept equal with each other, hence the effectiveness of optimization was limited.

To propose a universal optimization method, we began with a formal model for encoding an X-ray spectrum $S(E)$ into experimental channel $D_{i}$ for an N-channel FSS. The mathematical definitions and notations used in this paper are written as that Fehl used\cite {fehl_characterization_2010,fehl_characterization_2010-1}. $D_{i}$ can be written as
\begin{equation}
  \label{eq:ResponseModel}
  D_{i} = \int_0^{E_{\rm MAX}}R_{i}(E)S(E)dE+\epsilon_{i} = d_{i} + \epsilon_{i} (i=1,...N),
\end{equation}
where $D_{i}$ is the experimental channel data, $\{R_{i}(E)\}_{i=1}^{N}$ is the response function, $d_{i}$ is the noise-free channel data, and $\epsilon_{i}$ denotes the uncertainties and noise. Eq.(\ref{eq:ResponseModel}) can be written in vector-matrix notation:
\begin{equation}
  \label{eq:ResponseModelVecMat}
  \mathbf{D} = \mathbb{R}\mathbf{S}+\Delta\mathbf{D} = \mathbf{d}+\Delta\mathbf{D}.
\end{equation}

The errors affecting the unfolding accuracy comprise two parts. The first is perturbation $\Delta\mathbf{D}$ superimposed on $\mathbf{D}$ in the data-gathering process, e.g., statistical uncertainty, signal noise, digitization errors, uncorrected signal baseline, etc. The second is the bias between the experimentally measured or simulated RM and the real RM, $\mathbb{R} \rightarrow \mathbb{R}+\Delta\mathbb{R}$. $\Delta\mathbb{R}$ may result from drifts and uncertainties of the fit parameters, e.g., the light yield and light collection efficiency calibrated in section \ref{Sec02_SubSec03}. Since $(\Delta\mathbb{R})\mathbf{S}$ is analogous to $\Delta\mathbf{D}$\cite{fehl_characterization_2010-1}, only $\Delta\mathbf{D}$ is discussed in the following text.

Solving Eq.(\ref{eq:ResponseModelVecMat}) is always an ill-posed problem, and many unfolding algorithms can be used, e.g., the regularization methods, least-squares spectrum adjustment, parameter estimation, iterative unfolding methods, and the maximum entropy principle, etc\cite{reginatto_overview_2010}. These unfolding algorithms can hardly provide mathematically rigorous or realistic estimates of the error propagation relation of the spectral unfolding error $\Delta\mathbf{S}$ and the data perturbation $\Delta\mathbf{D}$. But the error propagation relation of $\Delta\mathbf{S}$ and $\Delta\mathbf{D}$ is mainly determined by the response function $\mathbb{R}$. For example, an upper-bound measure to $\Delta\mathbf{S}$ can be estimated from $\Delta\mathbf{D}$ and the condition number of RM $\rm cond(\mathbb{R})$ as\cite{fehl_characterization_2010}
\begin{equation}
  \label{eq:UpperBoundOfError}
  ||\Delta\mathbf{S}||/||\mathbf{S}|| \leq {\rm cond(\mathbb{R})}||\Delta\mathbf{D}||/||\mathbf{D}||,
\end{equation}
where $||\mathbf{D}||=(\sum_{i=1}^N D_{i}^2)^{\frac{1}{2}}$. Reducing $\rm cond(\mathbb{R})$, e.g., alleviating the ill condition of the RM, can lower the upper-bound measure to $\Delta\mathbf{S}$, thus improving the spectrum measurement precision. Moreover, $\rm cond(\mathbb{R})$ is determined by the configuration of FSS only. Therefore $\rm cond(\mathbb{R})$ can be used as the figure of merit of FSS optimization, which is independent of the spectrum.

Another challenge for spectrometer optimization is the multi-parameter optimization problem. The FSS often comprises tens of filter and sensor layers, the thicknesses and density (determined by the material) of each layer could affect $\rm cond(\mathbb{R})$. Therefore, there are tens of parameters to be optimized. Since the practical layer thickness has limitations for both the filter and sensor, and there is a lack of derivatives and linearity in this problem, the genetic algorithm is an appropriate method for optimization.

Some realistic factors limit the selection range of optimization parameters. FSS has thinner filters and sensors close to the X-ray sources, and thicker filters and sensors far away from the X-ray sources generally, and this can promise that most of the low energy X-rays deposit energies in the layers close to the X-ray sources and most of the high energy X-rays deposit energies in the layers far away from the X-ray sources, i.e., the RM has smaller $\rm cond(\mathbb{R})$. Considering that the light collection efficiency of the thin scintillator would be low, which could lead to a large statistical error, PIN diodes are used as sensors in the layers close to the X-ray sources. The PIN diodes are the S3590-08 type with 300 $\upmu m$ depletion layer thickness manufactured by Hamamatsu. Considering the convenience of the customized production of scintillators and filters, the materials of scintillators are $\rm Gd_3Al_2Ga_3O_{12}(GAGG)$\cite{iwanowska_performance_2013} and the materials of filters are aluminum or copper. The number of layers is a tradeoff between the applicable energy range and complexity of the signal readout system, and a total of 15 layers of sensors, including six PIN diodes and nine scintillators, are chosen. Hence, the general configuration of the online FSS is shown in TABLE.\ref{tab:FSSSimpleConfig}. The number of layers represents the distance between the layer and the X-ray source from the near to the far, and the filter is closer to the source in each layer. Since there is always an aluminum film of 50 $\upmu$m used as the first layer for electromagnetic shielding, the parameters to be optimized are the thicknesses of filters in layers 2-15 and the thicknesses of scintillators in layers 7-15.

\begin{table}[htbp]
	\centering
	\caption{The general configuration of the online FSS}
    \begin{tabular}{cccccc}
    \hline
      Layer & Filter & Sensor & Layer & Filter & Sensor\\ 
      \hline
      1 & Al & S3590-08 & 9 & Cu & GAGG \\
      2 & Al & S3590-08 & 10 & Cu & GAGG \\
      3 & Al & S3590-08 & 11 & Cu & GAGG \\
      4 & Al & S3590-08 & 12 & Cu & GAGG \\
      5 & Al & S3590-08 & 13 & Cu & GAGG \\
      6 & Al & S3590-08 & 14 & Cu & GAGG \\
      7 & Cu & GAGG & 15 & Cu & GAGG \\
      8 & Cu & GAGG &   &   &   \\
      \hline
  \end{tabular}
	\label{tab:FSSSimpleConfig}
\end{table}

\begin{figure*}[!htb]
	\includegraphics[width=0.95\textwidth]{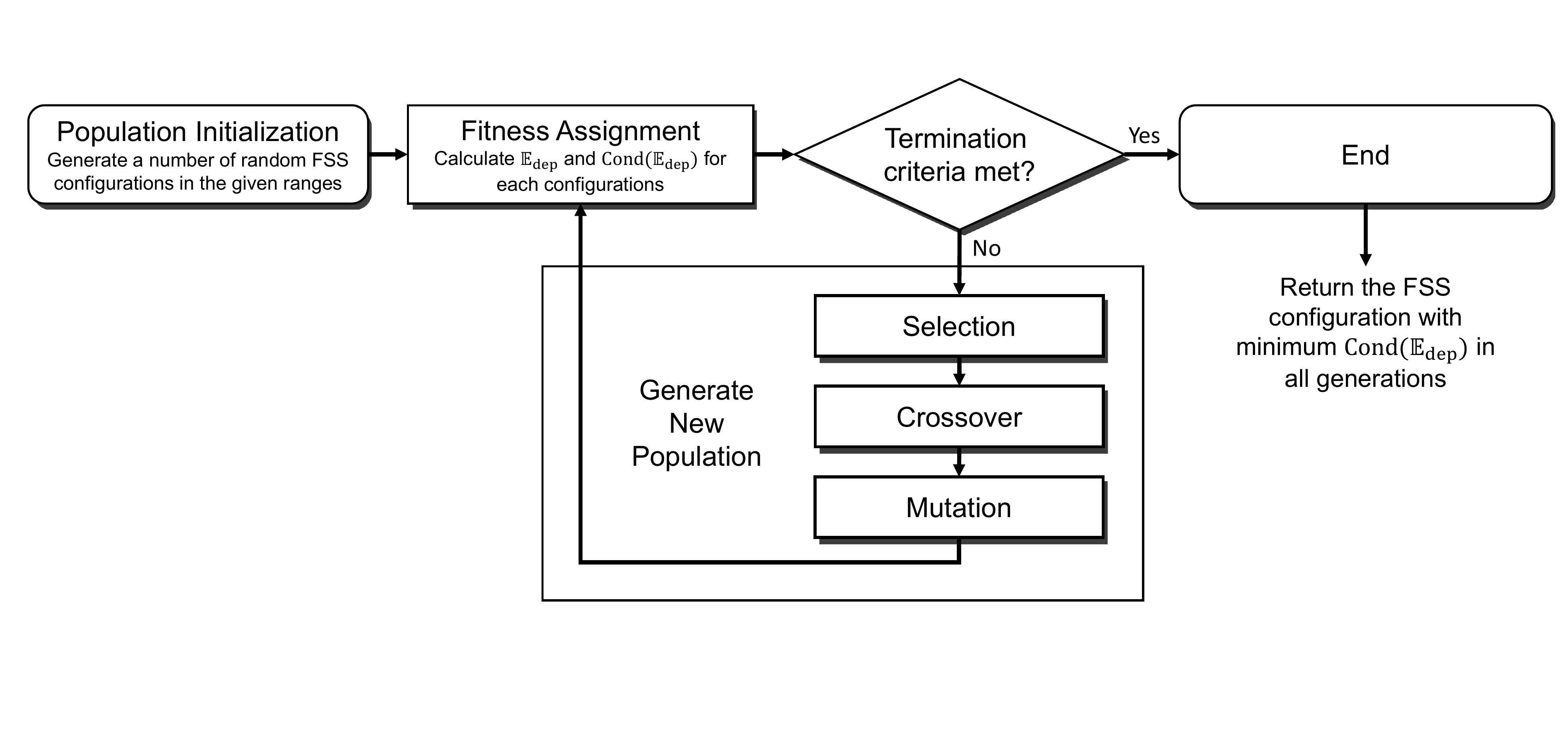}
	\caption{The optimization process of the online FSS using the genetic algorithm}
\label{fig:GAOptimizationProcedure}
\end{figure*}

The key point of the genetic algorithm implementation is the method of fitness assignment. Since there are 2000 individuals in the population and 300 times of iteration, $6\times 10^5$ times of $\rm cond(\mathbb{R})$ calculation, i.e., RM calculation, are needed approximately. The accurate RM can be calculated using Monte Carlo method\cite{jeon_broadband_2015,behm_spectrometer_2018,rusby_novel_2018}, and consumption of huge computational power would be required. To reduce the computational power requirement, a simple RM calculation model is proposed using the X-ray mass energy absorption coefficients and mass attenuation coefficient. First, the expected energy deposition of a single X-ray with energy $E$ in $i$th sensor $E_{{\rm dep}, i}(E)$ can be calculated as
\begin{equation}
  E_{{\rm dep},i}(E) = (1-e^{-\mu_{en,i}(E){\rm s}_{i}})\cdot\prod_{j=1}^{i-1}e^{-\mu_{j}(E){\rm s}_{j}}\cdot\prod_{j=1}^{i}e^{-\mu_{j}(E){\rm f}_{j}}E,
  \label{eq:SimpleEnergyDepositionCalModel}
\end{equation}
in which $\mu_{en,i}(E)$ is the X-ray mass energy-absorption coefficient of $i$th sensor, $\mu_{j}(E)$ is the X-ray mass attenuation coefficient of $j$th sensor or filter, ${\rm s}_{j}$ and ${\rm f}_{j}$ is the thickness of $j$th sensor and filter respectively. Then the RM can be written as
\begin{equation}
  R_{i}(E) = E_{{\rm dep},i}(E)\cdot ESC_{i}.
  \label{eq:SimpleRMCalModel}
\end{equation}
$ESC_{i}$ is the energy-signal coefficients of $i$th channel. For PIN diode
\begin{equation}
  ESC_{i}=1/3.62\times q\times \rho_{i},
  \label{eq:ESCofSiPIN}
\end{equation}
where $3.62$ eV is the pair creation energy of silicon\cite{owens_compound_2004}, $q$ is the elementary charge, $\rho$ is the gain of the electronics in ohm. For GAGG
\begin{equation}
  ESC_{i}=(LY\times LCE\times QE)_{i}\times q\times \rho_{i},
  \label{eq:ESCofGAGG}
\end{equation}
where $LY$ is the light yield of GAGG, $LCE$ is the light collection efficiency, and $QE$ is the quantum efficiency of the photon detector. Omitting the nonlinearity of scintillator light yield and electronics, $ESC_{i}$ is independent of X-ray energy $E$. Thus the practical response equation to be solved can be derived from Eq.(\ref{eq:ResponseModel}) and Eq.{\ref{eq:SimpleRMCalModel}}:
\begin{equation}
  \label{eq:ResponseFunc}
  \mathbf{D}/\mathbf{ESC} = \mathbb{E_{\rm dep}}\mathbf{S_{unfold}}.
\end{equation}
in which $\mathbf{S_{unfold}}$ is the spectral unfolding. There is an error $\Delta\mathbf{S}$ between $\mathbf{S_{unfold}}$ and the real X-ray spectrum $\mathbf{S}$ due to the perturbation $\Delta\mathbf{D}$ in $\mathbf{D}$. In Eq.(\ref{eq:ResponseFunc}), $\mathbb{E_{\rm dep}}$ is equivalent to the response matrix $\mathbb{R}$, and can be acquired from Eq.(\ref{eq:SimpleEnergyDepositionCalModel}). Hence, the optimization process to reduce $\rm cond(\mathbb{E_{\rm dep}})$ is shown in FIG.\ref{fig:GAOptimizationProcedure}.

The genetic algorithm is implemented using the Sheffield Genetic Algorithm Toolbox\cite{chipperfield_genetic_1994}. There are 2000 individuals in the population, and each individual contains 14 filter thicknesses and 9 scintillator thicknesses. These thicknesses are floating-point numbers and encoded by bit strings to create the chromosomes, i.e., each individual contains 24 binary chromosomes. Considering the limitation of practical layer thickness, the lower limits of 0.01 mm and upper limits of 50 mm are set for the filters and the lower limits of 2 mm and upper limits of 50 mm are set for the scintillators, and 20 bits in one binary chromosome are used to ensure enough precision. All the filter thicknesses will be set to 0 while the thicknesses lower than 0.1 mm in iterations. The generation gap, mutation probability, and probability of crossover are 0.9, 0.01, and 0.7 respectively. The termination criteria is the iteration time reaching 300.

\begin{figure}[!htb]
	\includegraphics[width=0.45\textwidth]{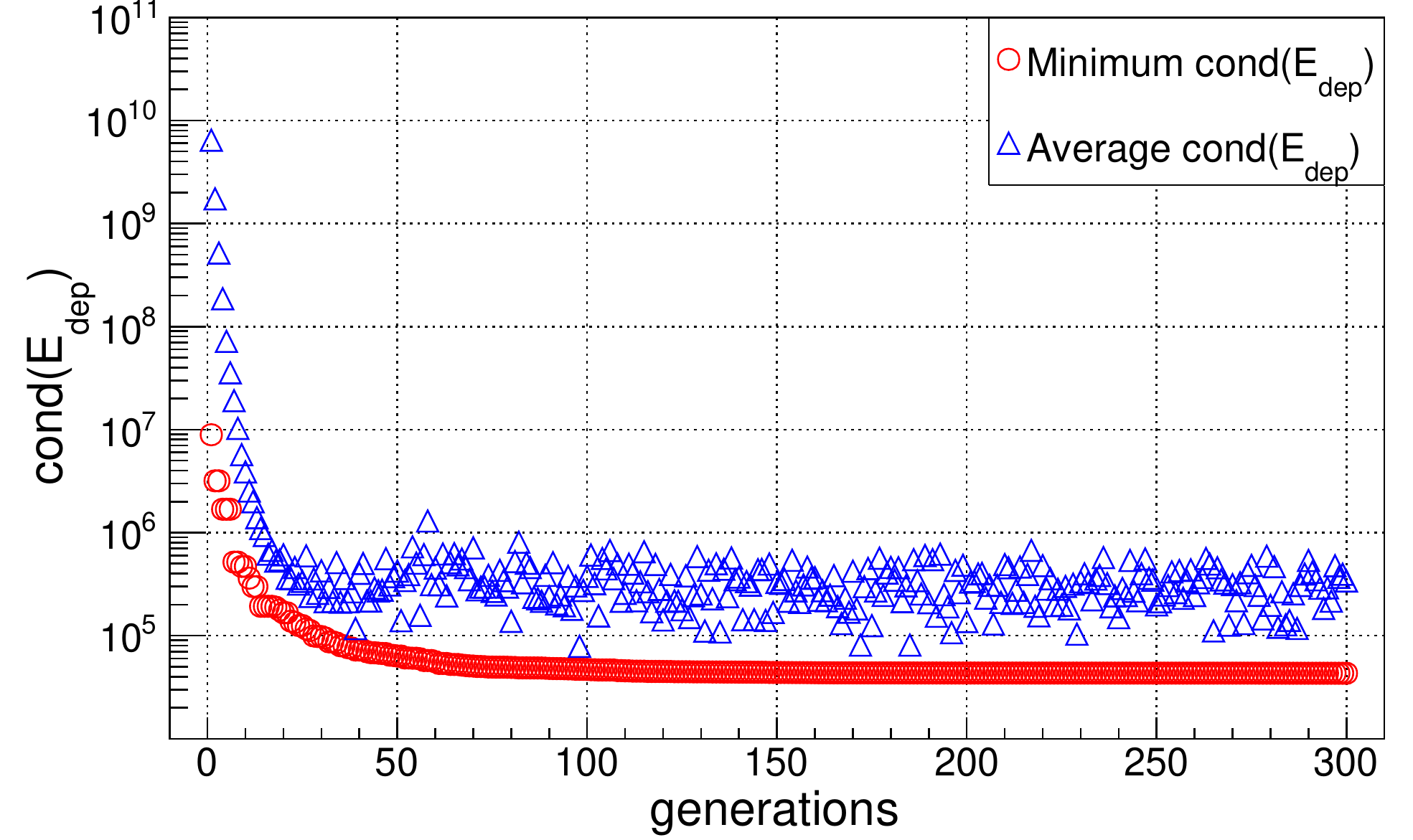}
	\caption{The evolution of $\rm Cond(\mathbb{E_{\rm dep}})$ in the genetic algorithm iterations}
\label{fig:OptimizationProcess}
\end{figure}

The minimum and average $\rm Cond(\mathbb{E_{\rm dep}})$ in the iterations is shown in FIG.\ref{fig:OptimizationProcess}. The optimal $\rm Cond(\mathbb{E_{\rm dep}})$ is $4\times10^{4}$ approximately, which has been significantly reduced compared with the unoptimized online FSS configuration, e.g., the CsI FSS reported previously\cite{behm_spectrometer_2018} with $\rm Cond(\mathbb{E_{\rm dep}})$ of $2\times10^{12}$ (also 15 channels are taken into account). The optimal configuration obtained in iterations is listed in TABLE.\ref{tab:FSSOptimalConfig}. The 0.01 mm accuracy is retained since the machining tolerance of GAGG crystal is 0.02 mm.

\begin{table}[htbp]
	\centering
	\caption{The optimized configuration of the online FSS}
    \begin{tabular}{ccccc}
    \hline
      Layer & Filter & \makecell[c]{Thickness \\ (mm) }& Sensor & \makecell[c]{Thickness \\ (mm) } \\
      \hline
       1 & Al & 0.05 & \multicolumn{2}{c}{S3590-08} \\
       2 & Al & 0 & \multicolumn{2}{c}{S3590-08}  \\
       3 & Al & 0 & \multicolumn{2}{c}{S3590-08}  \\
       4 & Al & 2.82 & \multicolumn{2}{c}{S3590-08} \\
       5 & Al & 9.95 & \multicolumn{2}{c}{S3590-08} \\
       6 & Al & 24.34 & \multicolumn{2}{c}{S3590-08} \\
       7 & Cu & 3.53 & GAGG & 2 \\
       8 & Cu & 10.07 & GAGG & 2 \\
       9 & Cu & 4.73 & GAGG & 4.44 \\
       10 & Cu & 6.97 & GAGG & 13.04 \\
       11 & Cu & 0.57 & GAGG & 12.15 \\
       12 & Cu & 13 & GAGG & 33.74 \\
       13 & Cu & 1.14 & GAGG & 21.65 \\
       14 & Cu & 18.24 & GAGG & 33.2 \\
       15 & Cu & 34.52 & GAGG & 50 \\
       \hline
  \end{tabular}
	\label{tab:FSSOptimalConfig}
\end{table}

\subsection{Unfolding spectrometer response}
\label{Sec02_SubSec02}

For testing the unfolding procedure and assessing the accuracy of unfolded radiation spectra, numerical experiments were conducted. From Eq.(\ref{eq:ResponseModelVecMat}) and Eq.(\ref{eq:SimpleRMCalModel}), the expected energy deposition $\mathbb{E_{\rm dep}}$, X-ray spectrum $\mathbf{S}$, energy-signal coefficients $\mathbf{ESC}$, and perturbation $\Delta\mathbf{D}$ were calculated and modeled to simulate the channel data.

\begin{figure}[!htb]
	\includegraphics[width=0.45\textwidth]{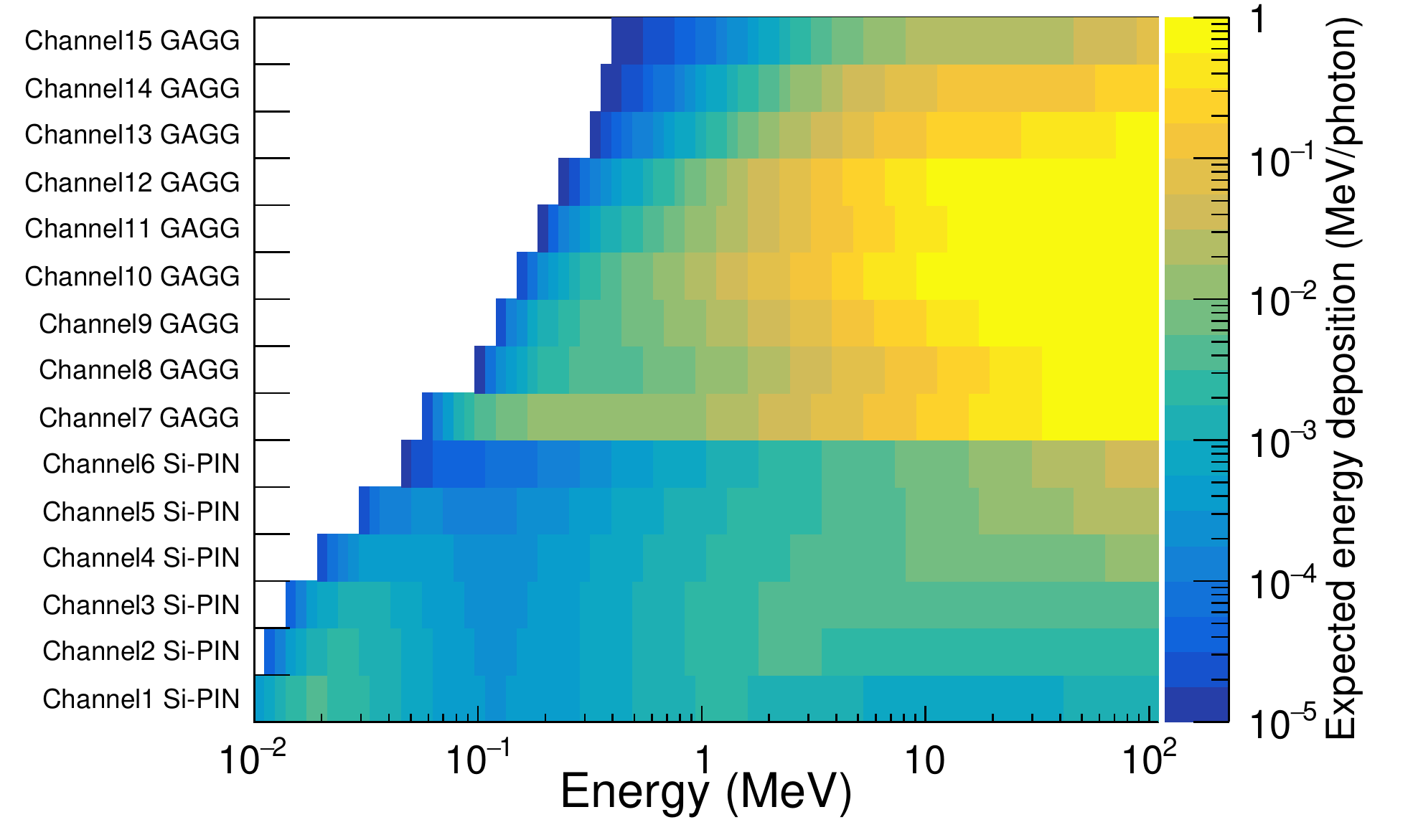}
	\caption{The response matrix expressed as $\mathbb{E_{\rm dep}}$ obtained by Monte Carlo simulations}
\label{fig:RMMonteCarlo}
\end{figure}

First, the optimized online FSS was modeled by using Monte Carlo code GEANT4\cite{agostinelli_geant4simulation_2003} to obtain the accurate $\mathbb{E_{\rm dep}}$. The simulation model included the filters, sensors, scintillation photon detector, as well as PIN diode carrier board, collimator, shielding, and mechanical parts. The incident X-rays were a pencil-like beam distributing uniformly and randomly in a circle with a diameter of 6 mm, which was the diameter of the collimator. The energy range of incident X-rays was 10 keV to 200 MeV. In each energy bin, X-ray energies were uniformly and randomly sampled, and $10^7$ X-rays were simulated. The energy deposition curves of these X-ray beams impinging on the online FSS, i.e., $\mathbb{E_{\rm dep}}$, is shown in FIG.\ref{fig:RMMonteCarlo}.

The X-ray spectrum $\mathbf{S}$ was modeled by
\begin{equation}
    S(E)=S_{\rm Beta}(E)+S_{\rm Brems}(E)+S_{\rm BG}(E),
    \label{eq:XRaySpectrum}
\end{equation}
where $S_{\rm Beta}(E)=AE^{\alpha}{\rm exp}(-\beta E)$ is the on-axis betatron component, $S_{\rm Brems}=\frac{B}{E}(\nu-{\rm ln}(E))$ is the on-axis bremsstrahlung component, and $S_{\rm BG}(E)=C{\rm exp}(-\eta E)$ is the background component\cite{schumaker_measurements_2014,jeon_broadband_2015}, and the unit of $E$ is MeV. The parameters $A$, $\alpha$, $\beta$, $B$, $\nu$, $C$, and $\eta$ define the curve shapes and fluence ratios of the three components, which were set to 12.47, 0.768, 7.52, $4.09\times 10^{-4}$, 3.88, 0.01, and 1 respectively. In this circumstance, the critical energy and peak energy of betatron radiation are 0.25 MeV and 0.1 MeV, and the fluence ratio of the background component is $7\%$. Assuming the energy of X-rays ranged from 10 keV to 10 MeV and the total photon number was $10^{5}$, the simulated energy depositions in the online FSS channels were obtained by multiplying the expected energy deposition $\mathbb{E_{\rm dep}}$ and the photon numbers in each energy bin $S(E){\rm d}E$.

The $\mathbf{ESC}$ can be estimated from Eq.(\ref{eq:ESCofSiPIN}) and Eq.(\ref{eq:ESCofGAGG}). The GAGG crystals used are the GAGG-HL type manufactured by EPIC CRYSTAL Co. Ltd. with a light yield $LY$ of 54000 photons/MeV. The light collection efficiency $LCE$ is $60\%$ approximately in the typical package and optical coupling conditions\cite{wen_compact_2021}. The maximum emission wavelength for GAGG is at around 530 nm\cite{iwanowska_performance_2013}, and the quantum efficiency $QE$ of PIN diodes, e.g., S3590-08, is $90\%$ approximately at the wavelength of 530 nm. Since the $\mathbf{ESC}$ could vary due to the different scintillators, packages, and optical couplings, it is further calibrated experimentally as described in Section \ref{Sec02_SubSec03}. The gains of FEE $\rho$ are the multiplying of the gains of TIAs and main amplifiers, which are determined to make the signal amplitudes conform to the dynamic range of each stage. The detailed discussion of $\rho$ can be found in Section \ref{Sec03_SubSec02} and the values are listed in TABLE.\ref{tab:FEEAmpRate}.

The perturbation $\Delta\mathbf{D}$ is mainly contributed by the statistical uncertainty and the electronic noise. The statistical uncertainty in PIN diode channels can be estimated by $\sigma(E)_{st}=2.355\sqrt{F\cdot E\cdot W}$, where $F=0.12$ is the Fano factor, $W$=3.62 eV is the energy for the formation of a charge carrier pair. The statistical uncertainties in GAGG channels are estimated from the previous calibration experiment conducted in our laboratory using radioactive sources\cite{wen_compact_2021}, i.e., $\sigma(E)_{st}=0.09\cdot 662\times 10^{3}\cdot\sqrt{\frac{662\times 10^{3}}{E}}$. The electronic noise is hard to be modeled and quantitatively estimated, an electronic noise $\sigma(E)_{el} = 0.01\times D_i$ is applied to all the channels, where $D_i$ means the noise-free channel data. It is practically straightforward to achieve an electronic system with this noise level.

\begin{figure}[!htb]
	\includegraphics[width=0.45\textwidth]{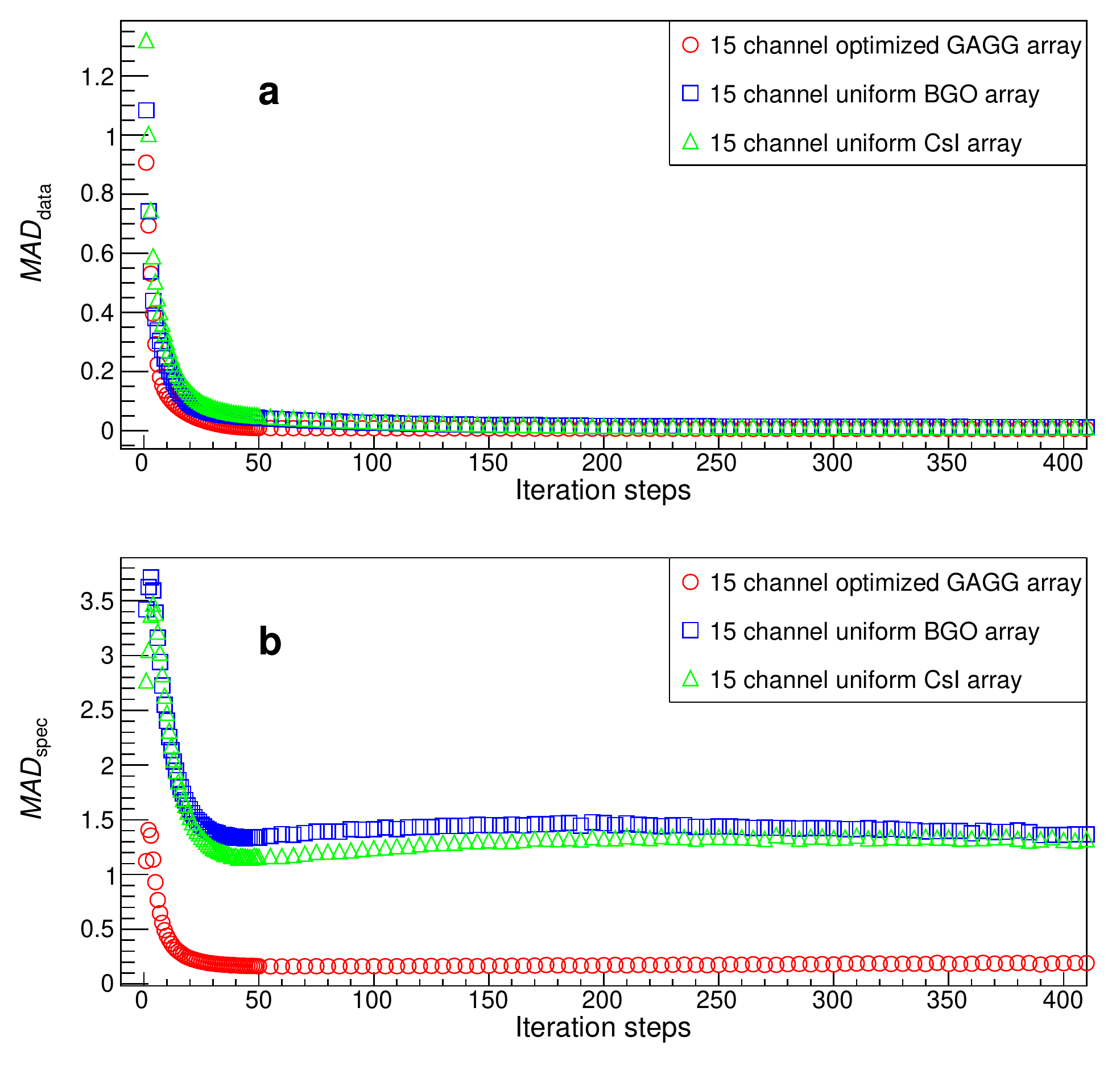}
	\caption{Converge curve (a) and unfolding error (b) for three kinds of online FSS}
\label{fig:ErrorComparison}
\end{figure}

\begin{figure}[!htb]
	\includegraphics[width=0.45\textwidth]{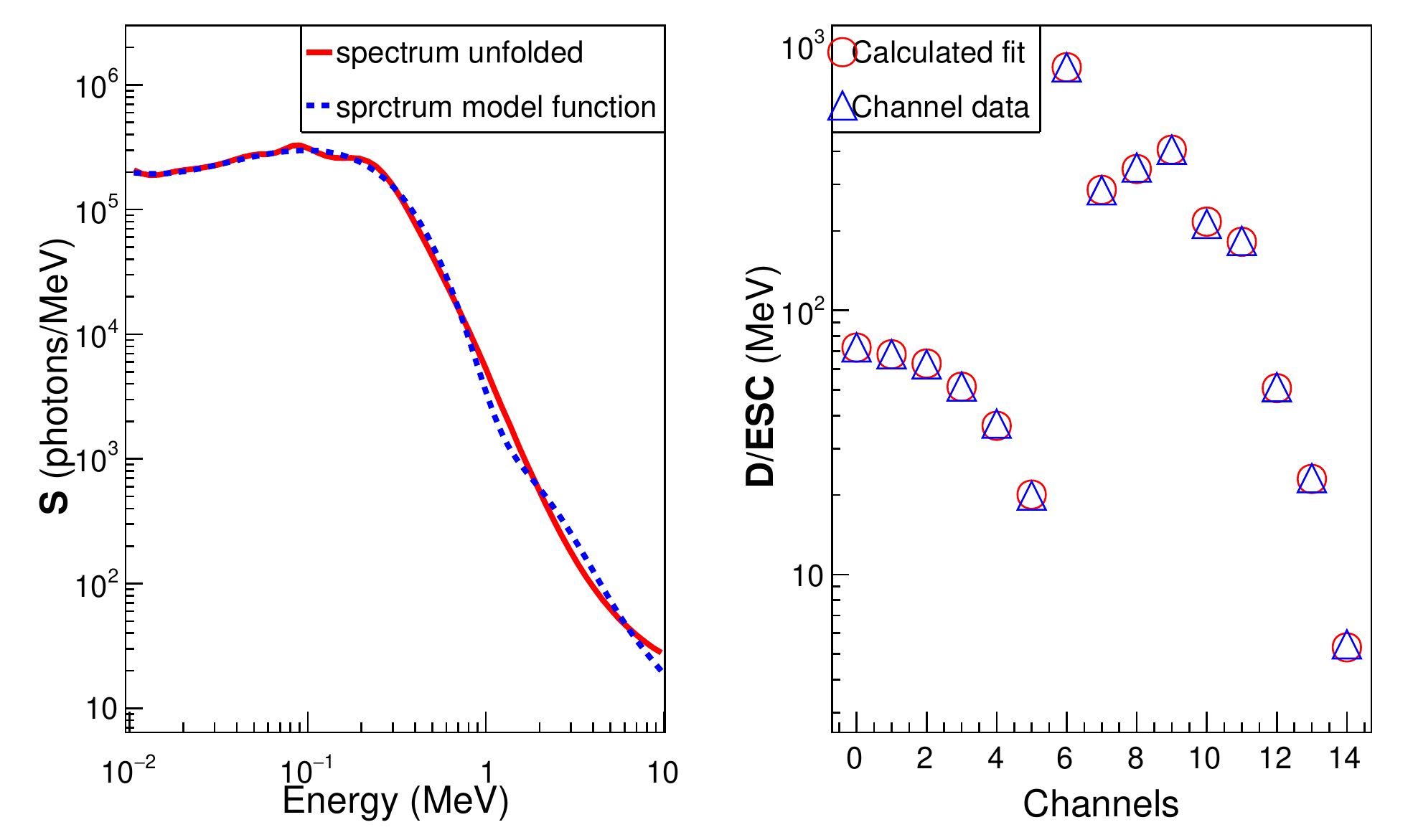}
	\caption{Fitting data (right) plotted with corresponding spectra (left)}
\label{fig:SpectrumUnfoldAndResponse}
\end{figure}

From the simulated channel data $\mathbf{D}_{\rm sim}$, the unfolded X-ray spectrum was further obtained by solving Eq.(\ref{eq:ResponseFunc}) using the expectation maximization method\cite{zhang_x-ray_2007}. The degree of convergence was indicated as the normalized mean absolute distance (MAD) between the fitting channel data and the simulated channel data
\begin{equation}
  MAD_{\rm data} = \frac{1}{N}\sum\frac{|D_{i,{\rm fit}}-D_{i,{\rm sim}}|}{D_{i,{\rm sim}}},
  \label{eq:MADData}
\end{equation}
where $N$ is the number of channels. With a constant function of one as the initial guess of $S_{\rm unfold}(E)dE$, the fitting error from a different number of iterations is plotted in Fig.\ref{fig:ErrorComparison}(a). For the optimized GAGG array (denoted as the red cross), the fitting error decreases rapidly to a low level at the first steps, then reaches <1\% after 50 steps. The divergence of the unfolded spectrum from the spectrum model function was also indicated as the normalized MAD between them
\begin{equation}
  MAD_{\rm spec} = \frac{1}{M}\sum\frac{|S_{\rm unfold}(E){\rm d}E-S(E){\rm d}E|}{S(E){\rm d}E},
  \label{eq:AveragedDeviation}
\end{equation}
where $M$ is the energy bin number. As Fig.\ref{fig:ErrorComparison}(b) shows, the divergence can reach a minimum of 16\% approximately in the energy range of 10 keV to 10 MeV when the fitting reaches convergence. The iteration step is chosen to be 100, which is large enough for observing the convergence property, a reasonable spectrum shape can be obtained and the response of the FSS is well fitted, as shown in FIG.\ref{fig:SpectrumUnfoldAndResponse}. It should be noted that the unfolding algorithms could significantly affect the unfolding results, and the expectation maximization method was chosen because it does not need many additional constraints and has a better universality.

For comparative analysis, the CsI array proposed by Behm\cite{behm_spectrometer_2018} and the BGO array proposed by Rusby\cite{rusby_novel_2018} were also simulated following the aforementioned steps, and the channel number and noise level remained the same with the GAGG array, i.e., 15 channels. The thicknesses of BGO crystals were 0.2 cm. We found that the condition number of the CsI array and BGO array are $2\times10^{12}$ and $1\times10^{12}$, and the minimum fitting errors are 115\% and 133\%, as shown in Fig.\ref{fig:ErrorComparison}. The unfolding error of the optimized GAGG array is significantly reduced compared with the CsI and BGO array of uniform thicknesses.

\subsection{Quantitative calibration of the spectrometer response}
\label{Sec02_SubSec03}

The accurate determination of RM is of vital importance to get the precise spectra unfolding. In Eq.(\ref{eq:SimpleRMCalModel}), $E_{{\rm dep},i}(E)$ can be obtained by Monte Carlo simulation precisely\cite{jeon_broadband_2015,behm_spectrometer_2018,rusby_novel_2018}. But the $\mathbf{ESC}$ must be experimentally calibrated for the scintillator channels since the light yields $LY$ and light collection efficiencies $LCE$ will vary depending on the scintillators, packages, and optical couplings and are difficult to be modeled and quantitatively calculated.

The most challenging thing in the experimental calibration campaign for $LY$ and $LCE$ is to determine the energy depositions in the scintillators. Behm et al. performed the experimental calibration by measuring the scintillation signal resulting from a bremsstrahlung interaction, where the bremsstrahlung X-ray beam was theoretically calculated in GEANT4 by simulating the collision of a typical electron beam with a 9 mm thick piece of lead. Then the energy depositions in the CsI array were further determined in GEANT4\cite{behm_spectrometer_2018}. Rusby et al. calibrated their online FSS by exposing the detector to radiation sources for a long period of time and integrating the camera images. The energy depositions in the period of time were also determined by the GEANT4 simulation\cite{rusby_novel_2018}.

In the optimized online FSS, the PIN diode S3590-08 is also used as the scintillation photon detector. Thanks to the 300 $\upmu$m depletion layer of S3590-08, the $LY\times LCE \times QE$ can be obtained by comparing the full-energy peaks of the radioactive sources, one of which directly irradiates the PIN diode S3590-08, and another of which irradiates the GAGG scintillator\cite{meng_design_2002}. The energy depositions in the scintillators, i.e., the gamma-ray lines of the radioactive sources, can be determined accurately, thus providing more reliable calibration results.

\begin{figure}[!htb]
	\includegraphics[width=0.45\textwidth]{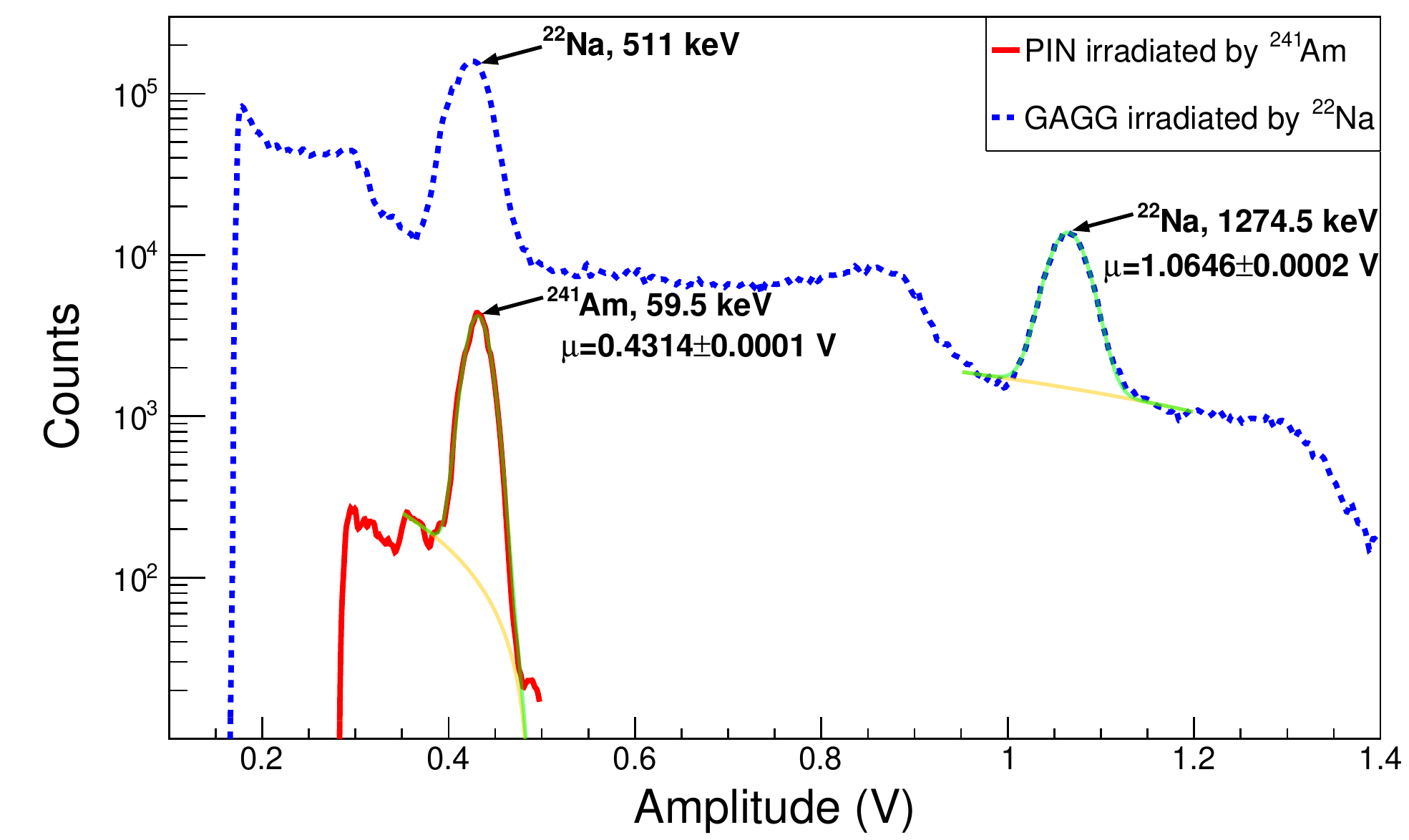}
	\caption{The spectrum obtained in the calibration experiment for one channel}
\label{fig:ExperimentalCalibrationResults}
\end{figure}

The $^{241}$Am with 59.5 keV line and $^{22}$Na with 1274.5 keV line are used in the calibration experiment. Charge sensitive pre-amplifiers, instead of TIA, were used to amplify the signals induced by radioactive sources. The spectrum measured during the experiment for one channel is shown in FIG.\ref{fig:ExperimentalCalibrationResults} as an example.

Since the readout electronics are identical for GAGG and PIN diodes, according to Eq.(\ref{eq:SimpleRMCalModel}), Eq.(\ref{eq:ESCofSiPIN}) and Eq.(\ref{eq:ESCofGAGG}), $LY\times LCE \times QE$ can be calculated by
\begin{equation}
  LY\times LCE \times QE = \frac{P_{\rm GAGG}E_{{\rm dep,PIN}}}{3.62[{\rm eV}]\cdot P_{\rm PIN}E_{{\rm dep,GAGG}}},
  \label{eq:CalibrationEquation}
\end{equation}
where $P_{\rm GAGG}$ and $P_{\rm PIN}$ are fitting peak positions in the experimental spectrum, $E_{{\rm dep,GAGG}}$ and $E_{{\rm dep,PIN}}$ are energy depositions in GAGG and PIN diodes, which are 1274.5 keV and 59.5 keV. For the spectrum shown in FIG.\ref{fig:ExperimentalCalibrationResults}, $P_{\rm GAGG}$ and $P_{\rm PIN}$ are $1.0646\pm 0.0002$ V and $0.4314\pm 0.0001$ V, therefore $LY\times LCE \times QE$ = $31826\pm 5$ electrons/MeV.

\section{Implementation}
\label{Sec03}
\begin{figure}[!htb]
	\includegraphics[width=0.45\textwidth]{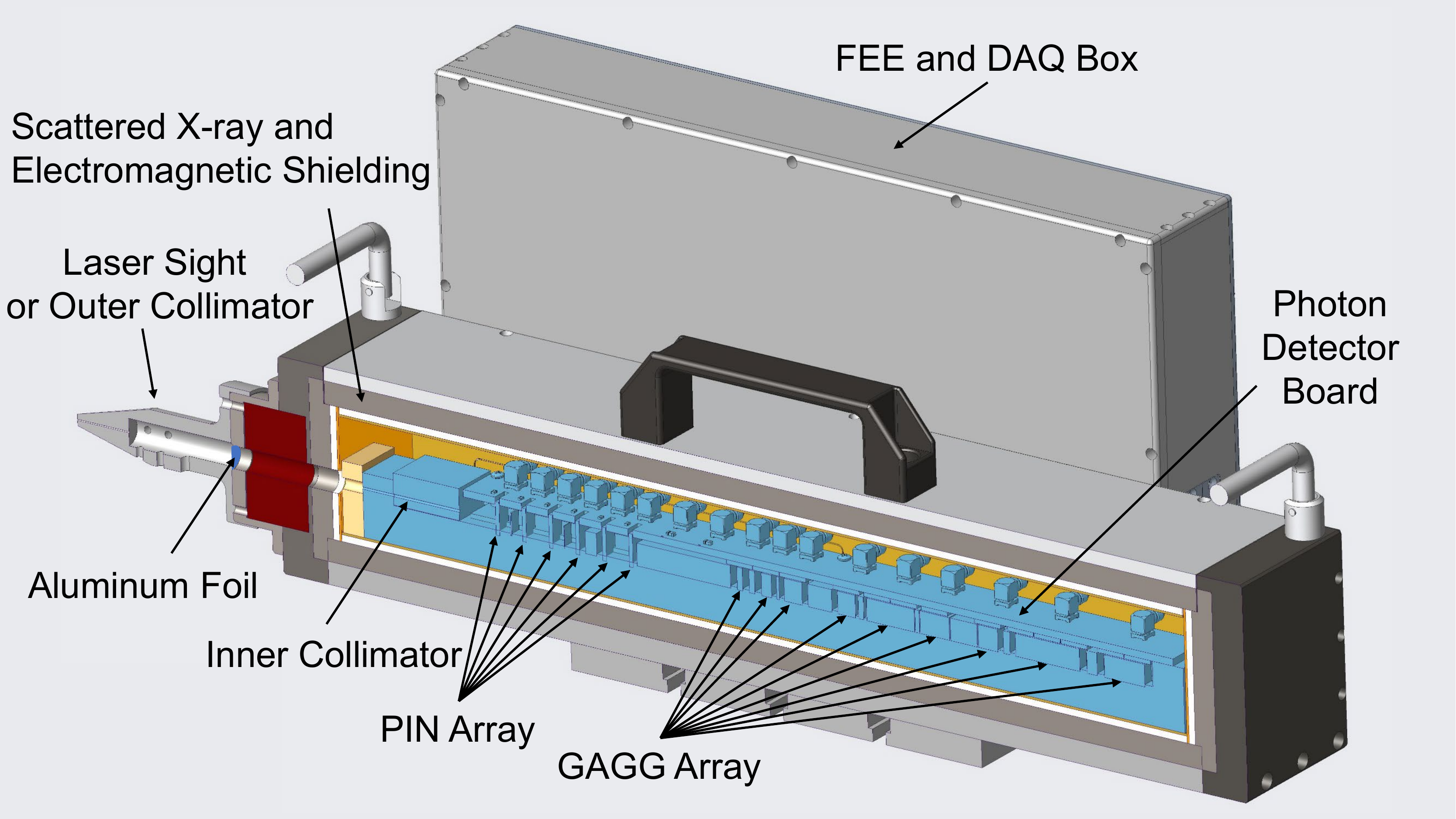}
	\caption{The section view of the online FSS prototype}
\label{fig:3DModelSectionView}
\end{figure}

An online FSS prototype was built and calibrated using the methods described above. A schematic of the online FSS prototype is shown in FIG.\ref{fig:3DModelSectionView}. The prototype comprises 6 channels of PIN diodes and 9 channels of GAGG scintillators as X-ray sensors, which are interleaved by the aluminum and copper filters. The collimator and laser sight are located in front of these PIN diode sensors, and the laser sight will be replaced by an outer collimator once the FSS is mounted. The scintillators are coupled with PIN diodes as the photon detectors. The signal output pins of the PIN diode are connected to the FEE board. The following DAQ board provides the capabilities of 16-channel waveform digitization and communication. And there is a power board providing the low noise $\pm 5$ V power and 70 V power for the FEE and PIN diode. The filter and X-ray sensors are shielded by a 1 cm thick layer of lead to avoid the interference of scattered X-rays, and there is a 1 cm diameter aperture allowing the X-ray incidence. The aperture is covered by a 50 $\upmu$m thick aluminum foil for electromagnetic shielding. The electronics are placed in an enclosed copper and aluminum box for electromagnetic shielding. The external interface only includes a 12 V power supply cable and a USB communication cable.

\subsection{Filter and sensor unit}
\label{Sec03_SubSec01}

GAGG scintillators were chosen due to their good mechanical characteristics, non-hygroscopic properties, high light yields (30-70 ph/keV), and fast decay times ($\sim$100 ns). The scintillators used have front faces of 1 cm $\times$ 1 cm and variable thicknesses. The filter and scintillator configuration is listed in TABLE.\ref{tab:FSSPrototypeConfig}. The configuration of the prototype was obtained by the original version of our optimization code without proper constraints of the thicknesses, thus different from the configuration listed in TABLE.\ref{tab:FSSOptimalConfig}. Nevertheless, the condition number of the prototype response matrix is $6\times 10^5$, which is close to the optimized condition number presented in section \ref{Sec02_SubSec01} (compared with the uniform thickness configuration), and the improved performances are expected.

\begin{table}[htbp]
	\centering
	\caption{The configuration of the online FSS prototype}
    \begin{tabular}{ccccc}
    \hline
      Layer & Filter & \makecell[c]{Thickness \\ (mm) }& Sensor & \makecell[c]{Thickness \\ (mm) } \\
      \hline
       1 & Al & 0.05 & \multicolumn{2}{c}{S3590-08} \\
       2 & Al & 0.59 & \multicolumn{2}{c}{S3590-08}  \\
       3 & Al & 0 & \multicolumn{2}{c}{S3590-08}  \\
       4 & Cu & 0.11 & \multicolumn{2}{c}{S3590-08} \\
       5 & Al & 0 & \multicolumn{2}{c}{S3590-08} \\
       6 & Al & 5.5 & \multicolumn{2}{c}{S3590-08} \\
       7 & Cu & 45.55 & GAGG & 2.6 \\
       8 & Cu & 2.63 & GAGG & 4.85 \\
       9 & Cu & 1.3 & GAGG & 8.67 \\
       10 & Cu & 11.59 & GAGG & 8.58 \\
       11 & Cu & 2.52 & GAGG & 23.26 \\
       12 & Cu & 2.06 & GAGG & 14.26 \\
       13 & Cu & 12.94 & GAGG & 9.49 \\
       14 & Cu & 3.16 & GAGG & 31.23 \\
       15 & Cu & 4.91 & GAGG & 19.58 \\
       \hline
  \end{tabular}
	\label{tab:FSSPrototypeConfig}
\end{table}

\begin{figure}[!htb]
	\includegraphics[width=0.45\textwidth]{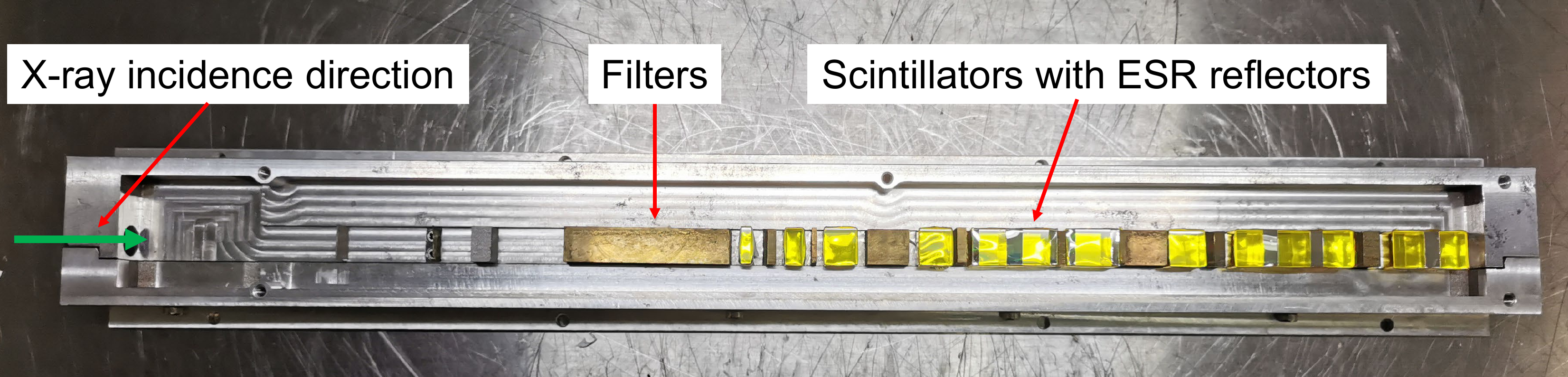}
	\caption{The filters and scintillators}
\label{fig:FiltersAndScintillatorsPhoto}
\end{figure}

The reflection layers of scintillators are enhanced specular reflectors (ESRs), 65 $\upmu$m polymer with high reflectance (>98\%) manufactured by 3M. The ESRs are cut into particular shapes allowing the scintillation light coupling from the side of the scintillators. The scintillators are fixed by an aluminum frame and interleaved by the filters, as shown in FIG.\ref{fig:FiltersAndScintillatorsPhoto}.

\begin{figure}[!htb]
	\includegraphics[width=0.45\textwidth]{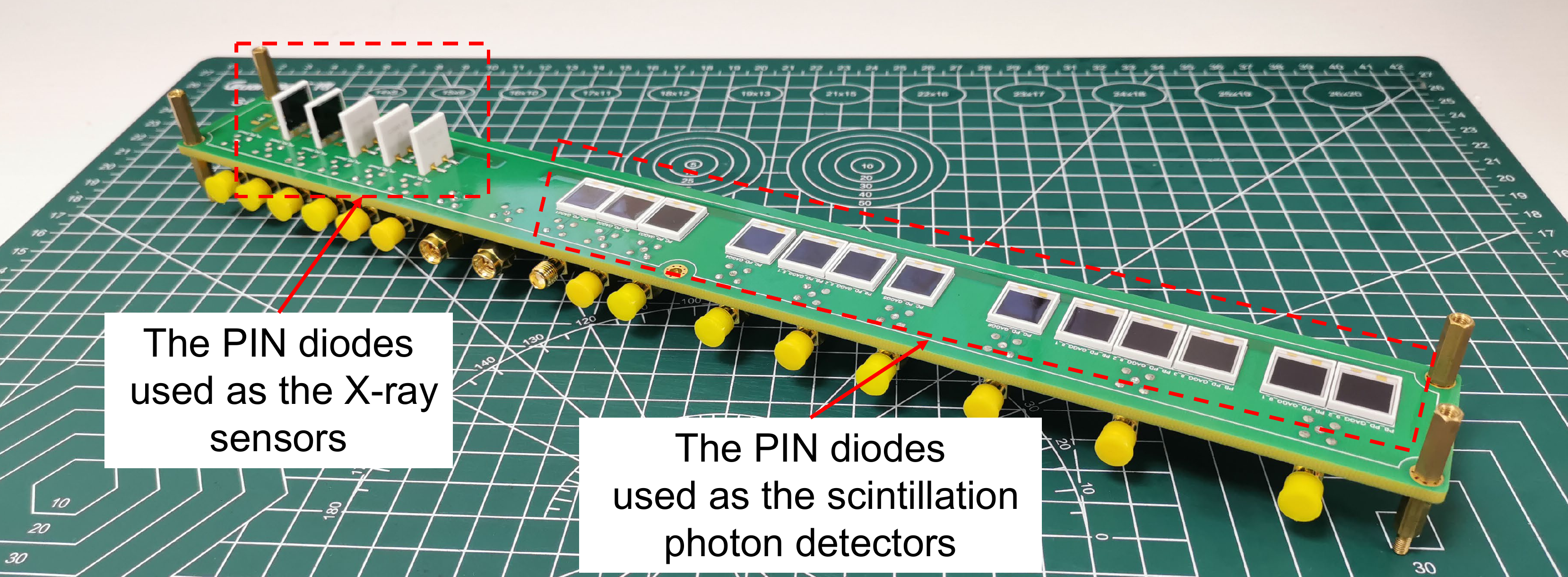}
	\caption{The PIN diodes and their carrier board}
\label{fig:PINDiodesPhoto}
\end{figure}

The scintillation light output will be bright when the scintillators irradiated by laser-induced brilliant X-ray pulses, e.g., >$5 \times 10^7$ scintillation photons ($10^3$ MeV $\times$ 54000 photons/MeV) per channel per pulse. The non-multiplying photon detector would be suitable for the scintillation light collection. To improve the light collection efficiency and the coupling stability, the PIN diode with a large sensitive area (1 cm $\times$ 1 cm) is used as the photon detector, and the PIN diodes are directly coupled with the scintillators using the optical grease. Since there are 6 channels of PIN diodes used as the X-ray sensors, all the PIN diodes are mounted on one PIN diode carrier board vertically or horizontally, as shown in FIG.\ref{fig:PINDiodesPhoto}.

\begin{figure}[!htb]
	\includegraphics[width=0.45\textwidth]{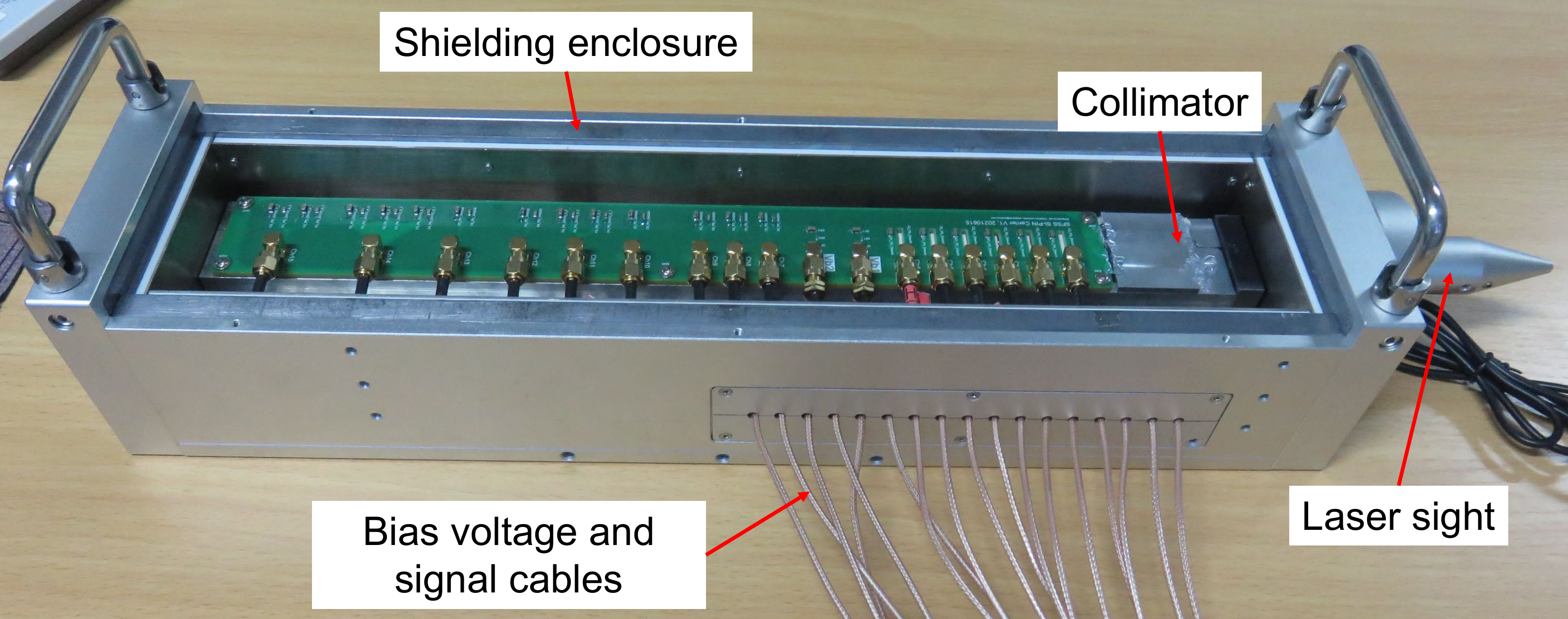}
	\caption{The filter and sensor unit}
\label{fig:FilterAndSensorUnit}
\end{figure}

The sensors, filters, and PIN diodes are placed in an X-ray and electromagnetic shielding enclosure, along with the collimator and laser sight making up the filter and sensor unit as shown in FIG.\ref{fig:FilterAndSensorUnit}. The bias voltages and electric signals are applied or elicited using the coaxial cables.

\begin{figure}[!htb]
	\includegraphics[width=0.45\textwidth]{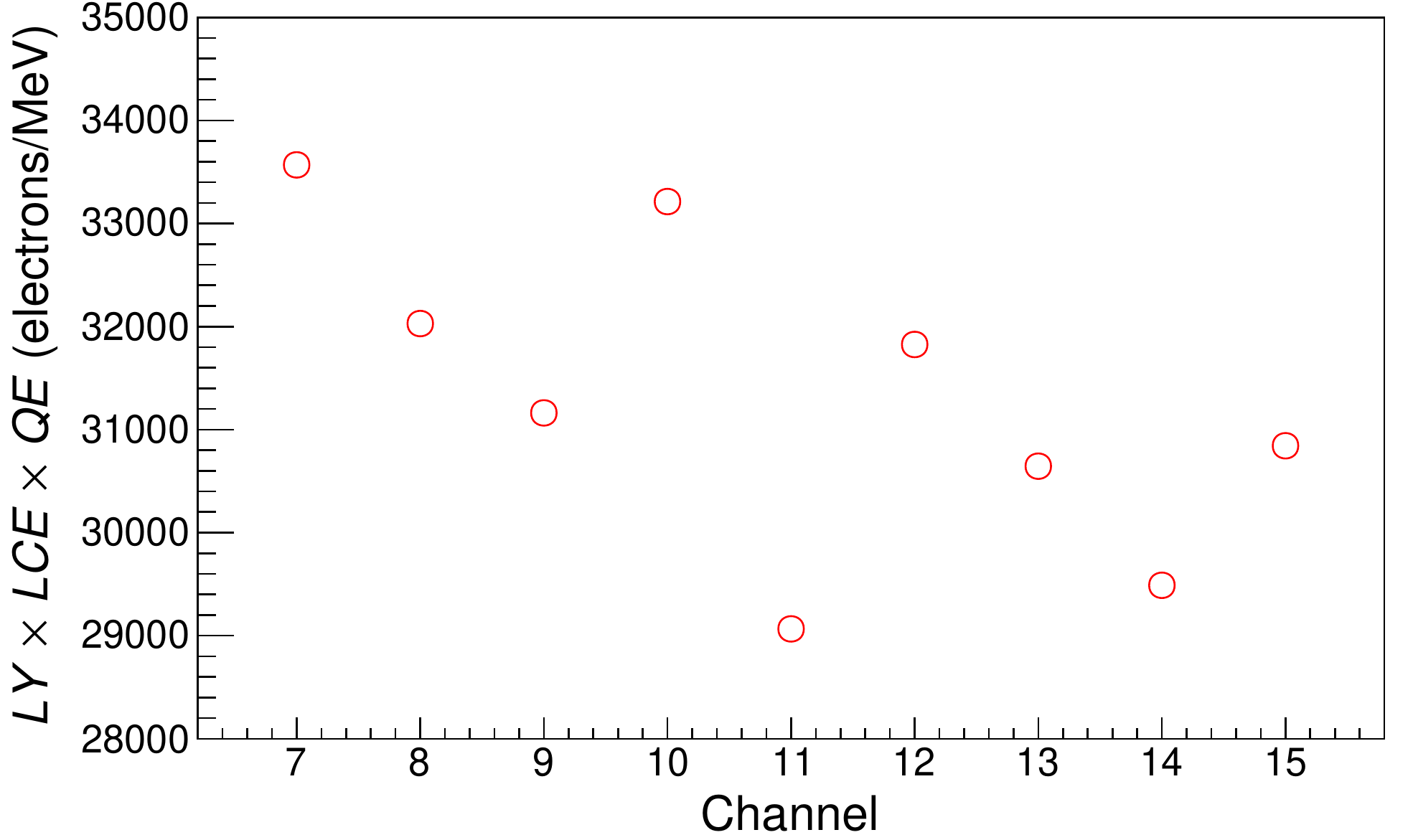}
	\caption{The experimentally calibrated $LY \times LCE \times QE$ for the 9 GAGG scintillation channels of the prototype}
\label{fig:LYLCEQECaliRes}
\end{figure}

Using the calibration method described in section \ref{Sec02_SubSec03}, the values of $LY\times LCE \times QE$ for 9 GAGG scintillation channels were experimentally calibrated as shown in FIG.\ref{fig:LYLCEQECaliRes}. The error bars are not visible in this figure due to the low peak fitting error.

\subsection{Electronics unit}
\label{Sec03_SubSec02}

The PIN diode signal is fed to a trans-impedance amplifier (TIA) via the alternating current (AC) coupling, and a standard high-speed amplifier OPA657 is adopted as the TIA amplifier. A 2000 ohms resistor connects the cathode of the PIN diode to the ground as the direct-current (DC) path, which can restrain the current of the PIN diode and increases the system reliability. The TIA is followed by the main amplifier, adjusting the voltage amplitude to match the DAQ dynamic range. In addition to 16-channel TIAs and main amplifiers, the FEE board includes a trigger circuit to generate the inner trigger signal for the DAQ.

Gains of the FEE circuit are determined to ensure the signal amplitude is in line with the dynamic range in each circuit stage. First, the energy depositions per pulse in the PIN diodes and scintillators are calculated using GEANT4. The X-ray source has a spectrum as Eq.(\ref{eq:XRaySpectrum}) modeled, photon fluence of $1\times 10^{11}$ photons/sr/pulse, and located at 1 meter away from the collimator of online FSS. Thus the number of X-ray photons passing through the collimating aperture is $2.8\times 10^{6}$ per pulse.

\begin{figure}[!htb]
	\includegraphics[width=0.45\textwidth]{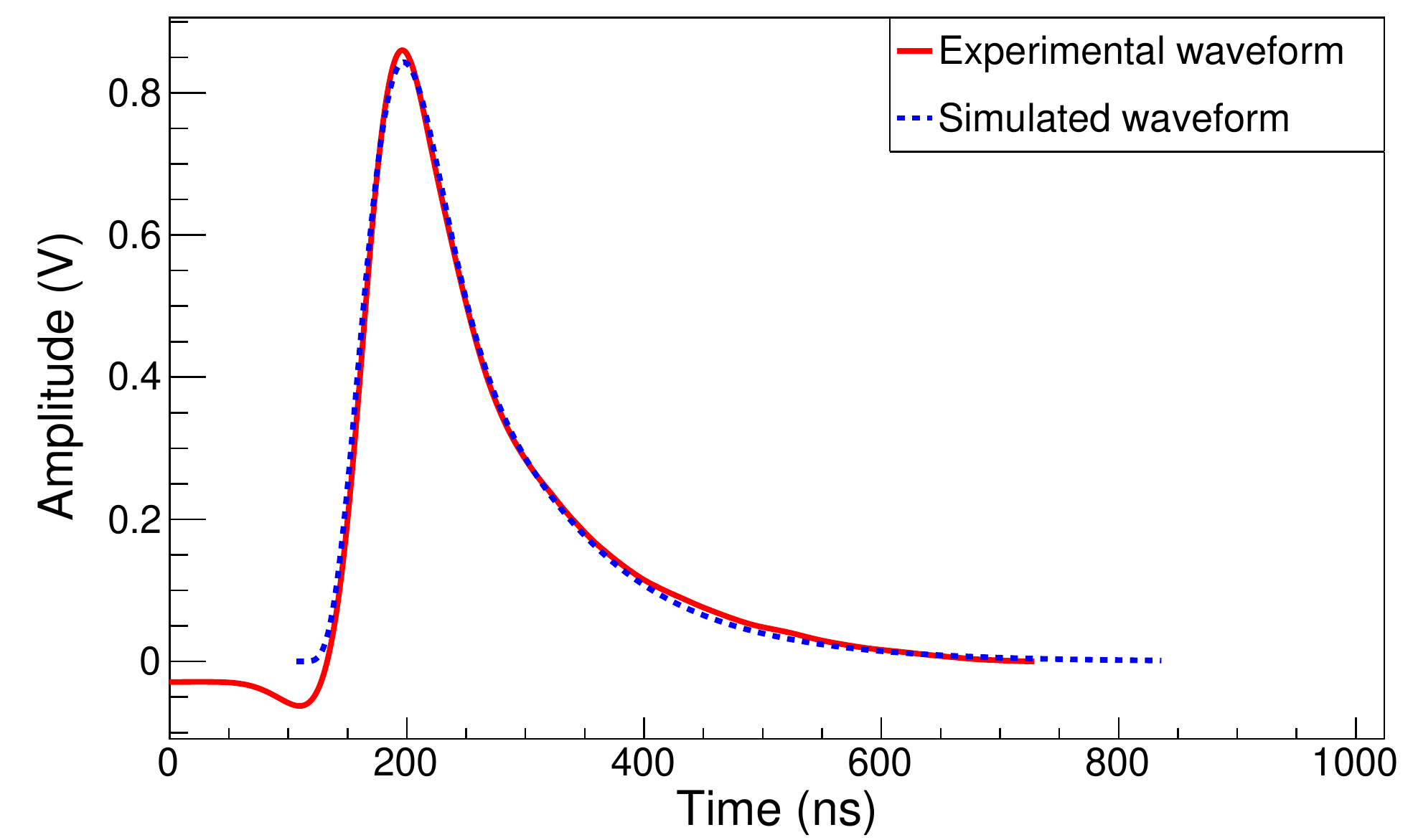}
	\caption{The simulated and experimental waveforms of the FEE output}
\label{fig:WaveformModelization}
\end{figure}

Then, the temporal waveforms are calculated using the transient response model, including the GAGG scintillation light, PIN diodes output current, and the output voltage of each circuit stage. The PIN diode is treated as a two-order Butterworth low-pass filter (LPF) with a -3 dB bandwidth of 40 MHz\cite{noauthor_s3590-08_nodate}. The transient response of the GAGG scintillator can be expressed as a single exponential decay signal with a decay time of 100 ns, and the TIA can also be treated as a two-order Butterworth LPF with -3 dB bandwidth of $\sqrt{GBP/(2\pi R_{\rm F}C_{\rm D})}$\cite{wen_compact_2021}, where $GBP$ is the gain bandwidth product of OPA657, $R_{\rm F}$ is the feedback resistance, and $C_{\rm D}$ is the input capacitance of the TIA. $C_{\rm D}$ is mainly contributed by the terminal capacitance of PIN Diodes and the stray capacitance of cables in the order of $\sim$100 pF. The main amplifier is designed to be an inverting two-order voltage-controlled voltage sources LPF with a -3 dB bandwidth of 10 MHz. The simulated waveform of the FEE output is plotted along with the experimental waveform measured in the laser-plasma experiment in Fig.\ref{fig:WaveformModelization}. The simulated waveform has preferably coherence with the experimental waveform, e.g., the relative difference of the peak amplitudes is less than 2\% when the integrals from 130 ns to 700 ns are equal (with the same energy depositions), thus validating the built transient response model.

Since the voltage output swing of TIA is 0-3 V and the DAQ input dynamic is 1 V, the gains of the main amplifier are determined to make the dynamic range of the DAQ fully used, which is 0.32. Then, using the transient response model, the gains of TIA are determined to make the peak amplitudes of 15-channel TIA outputs in the order of 500 millivolts, which are listed in TABLE.\ref{tab:FEEAmpRate}. The main purpose of this section is to describe the hardware design methods. As the X-ray fluence varies greatly in different laser-plasma experiments, the FEE is designed to be a plug-in module thus the gains can be determined following the methods and easily changed.

\begin{table}[htbp]
	\centering
	\caption{The gains of FEE}
  \begin{tabular}{c|ccc}
  \hline
   channels & 1-6 & 7-12 & 13-15 \\
   \hline
   $\rho$ in ohm & 20$\times$0.32 & 200$\times$0.32 & 1000$\times$0.32 \\
   \hline
  \end{tabular}
  \tabnote{The gains of TIAs and main amplifiers are listed on the left and right sides of multiplication signs respectively.}
	\label{tab:FEEAmpRate}
\end{table}

For reserving the temporal waveform information and further implementing the flexible digital signal processing technique, a commercial 16-channel high-speed analog-digital conversion module based on DRS4 chips is adopted as the DAQ system. The DAQ provides a 1 GHz sample rate, 1024 sample acquisition window (1.024 $\upmu$s at 1 GHz), and 30 MB/s transfer rate (USB2.0 protocol). Since the decay time of GAGG is 100 ns, a 1.024 $\upmu$s acquisition window can record the waveform completely. A higher repetition rate of hundred kHz should be allowed for the online FSS system, but the repetition rate is limited to 1 kHz approximately by the DAQ data transfer rate, i.e., 30 MB/s divided by "16 channel$\times$12 bit/sample$\times$1024 sample/channel/pulse" equals 1220 pulse/s. For future campaigns where higher repetition rates may be available, the online FSS system can be easily changed to a faster one by upgrading the data transfer protocol.

\section{Preliminary Experiment}
\label{Sec04}

\begin{figure}[!htb]
	\includegraphics[width=0.45\textwidth]{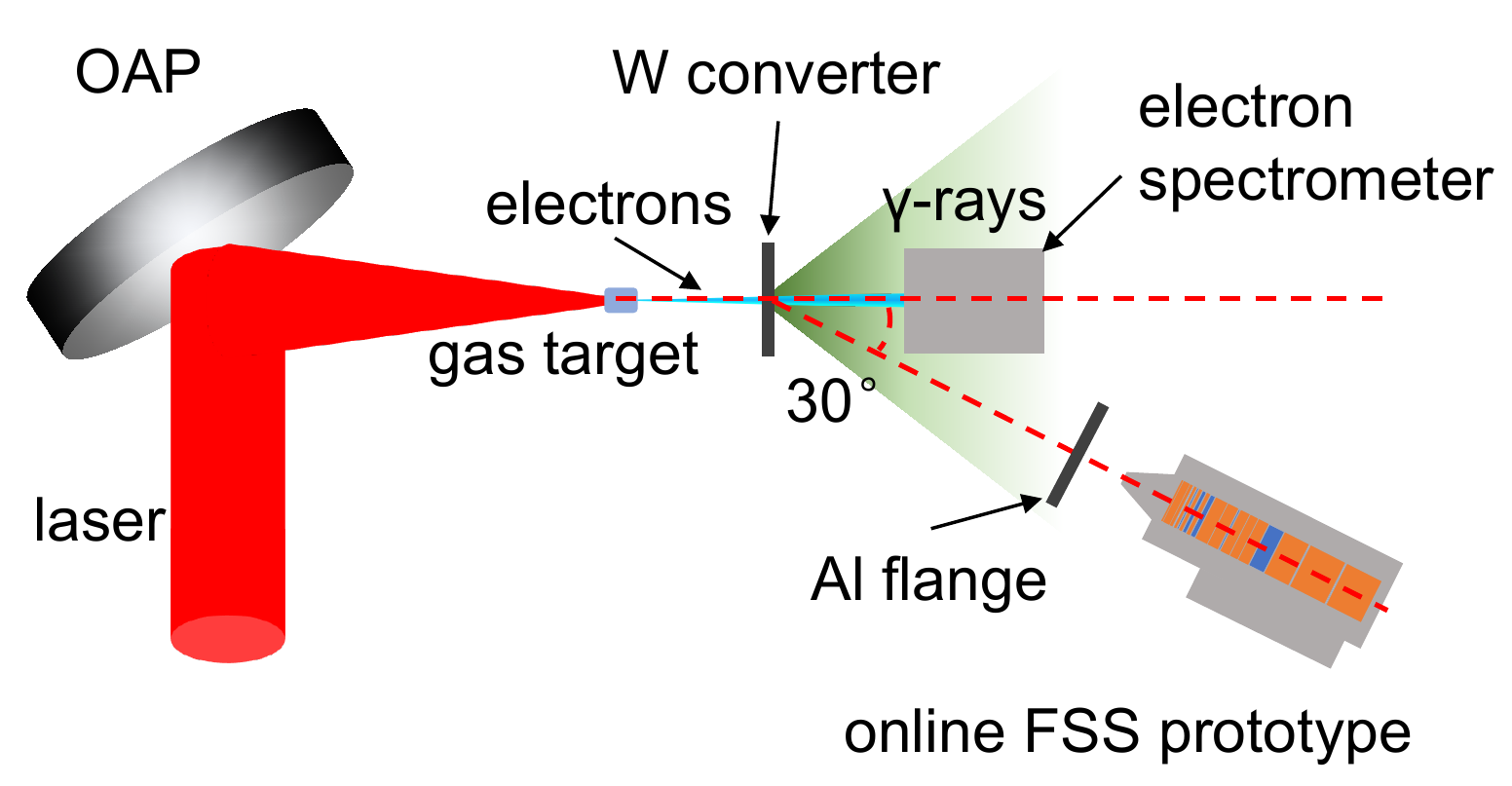}
	\caption{The experimental setup}
\label{fig:SILEXIIExperimentConfig}
\end{figure}

To test the functionality of this online FSS design, the prototype was deployed on a positron experiment, where strong bremsstrahlung was generated. The experiment was conducted at the SILEX-II laser, a complete optical parametric chirped-pulse amplification (OPCPA) laser operating at a wavelength of 800 nm\cite{hong_commissioning_2021}. The laser energy on targets is 27 $\pm$ 2.7 J (RMS), and the laser pulse duration is 34.5 $\pm$ 2.5 fs (RMS), thus the on-target laser power is around 1 PW. In the experiment, the laser interacted with a gas cell to generate high-energy electron beams via wakefield acceleration. The gas cell was 800 $\upmu$m in length and filled by 128 kPa nitrogen. The energetic electron beam then propagated through a converter material (2 mm tungsten) mounted 10 mm behind the gas cell to generate the bremsstrahlung gamma-ray beams. The online FSS prototype was located 1.5 m from the converter at an angle of 30$^\circ$ from the laser propagation direction, as shown in Fig.\ref{fig:SILEXIIExperimentConfig}.

\begin{figure}[!htb]
	\includegraphics[width=0.45\textwidth]{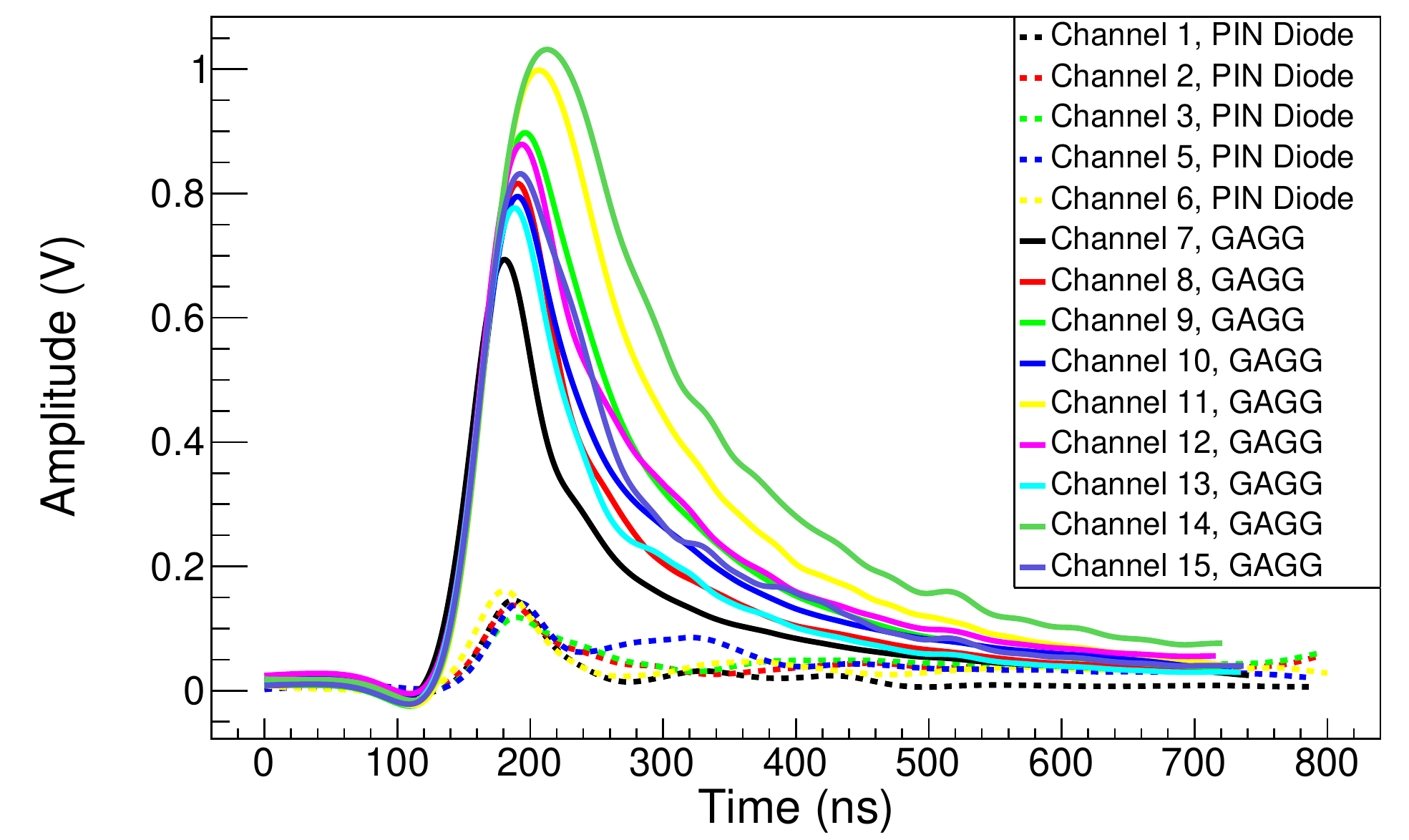}
	\caption{The measured waveforms}
\label{fig:ExpWaveform}
\end{figure}

Fig.\ref{fig:ExpWaveform} represents the measured waveforms in one shot. The disturbance induced by electromagnetic pulse could be observed in these waveforms, and the waveform of channel 4 was missed due to the loose contact. Nevertheless, the X-ray spectrum was still unfolded as a preliminary demonstration.

\begin{figure}[!htb]
	\includegraphics[width=0.45\textwidth]{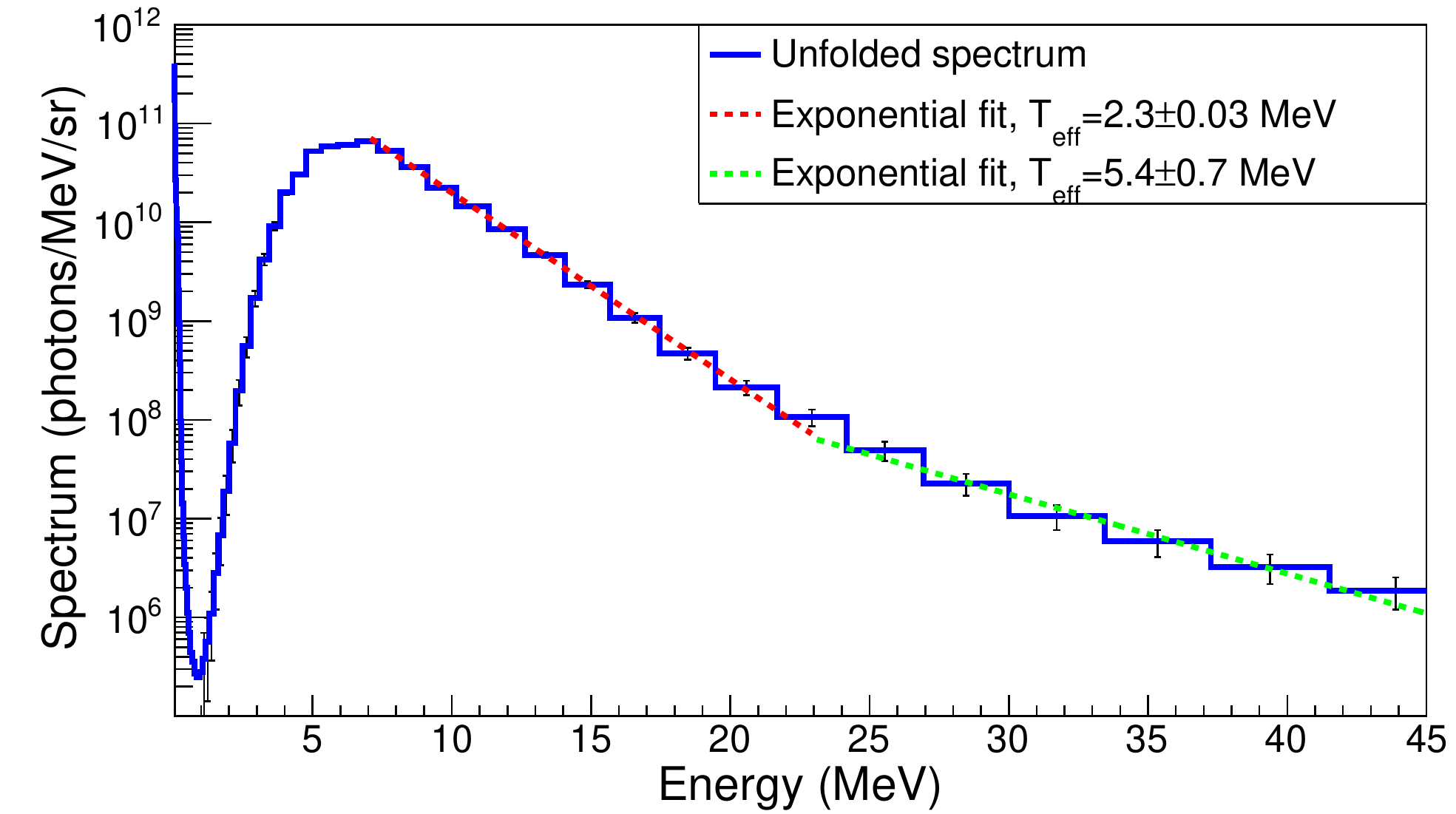}
	\caption{Spectrum unfolded in the preliminary experiment}
\label{fig:ExpSpectrumUnfolded}
\end{figure}

The waveforms were integrated from 130 ns to 700 ns for the GAGG channels and from 130 ns to 200 ns for the PIN diodes channels, as the experimental channel data $\mathbf{D}$. The $\mathbb{E_{\rm dep}}$ was obtained by Geant4 simulation and the $\mathbf{ESC}$ was calibrated and calculated according to Eq.(\ref{eq:ESCofSiPIN}) and Eq.(\ref{eq:ESCofGAGG}). The spectrum was obtained by solving Eq.(\ref{eq:ResponseFunc}) using the expectation maximization method with 100 steps of iteration, as shown in Fig.\ref{fig:ExpSpectrumUnfolded}, and the transmission of the Al flange had been taken into account. The error bars were given by adding random noises to the channel data and repeating the unfolding process $1\times 10^5$ times. Considering the unperfect electromagnetic shielding, we approximated the random noises as 5\% of the channel data. The spectral shape was consistent with a typical high power laser-induced bremsstrahlung spectrum\cite{zulick_high_2013}, and could be well fitted as a two effective temperature distribution with exponential effective temperatures of the form ${\rm d}N/{\rm d}E = a\times {\rm exp}(-E/T_{\rm eff})$. The lower effective temperature and the higher effective temperature were fitted between 7-23 MeV and 23-45 MeV respectively, yielding $T_{\rm eff,low}=2.3\pm0.03$ MeV and $T_{\rm eff,high}=5.4\pm0.7$ MeV.

\section{CONCLUSION}
\label{Sec05}

In conclusion, to alleviate the ill-posed system of FSS, an optimization method using the genetic algorithm is proposed. The condition number of RM is used as the figure of merit for optimization. We use not only high-density GAGG scintillators but also silicon PIN diodes, to improve the unfolding accuracy in tens of keV to tens of MeV energy bands. The Monto Carlo simulations show that the optimized FSS can provide a measurement of a typical spectrum from tens keV to tens MeV with a significantly improved accuracy.

We have implemented the optimized online FSS. PIN diodes are used as the scintillation photon detector to improve the light collection efficiency and coupling stability, thus further improving the unfolding accuracy. The usage of PIN diodes also promised the quantitative calibration of the light yields and light collection efficiencies. The electric signals from PIN diodes are amplified and digitalized by a homemade FEE module and commercial DAQ card. The repetition rate of the online FSS prototype is 1 kHz and can be further improved by increasing the data transfer rate of the DAQ card, e.g., using the 80 MB/s optical link interface or USB 3.0 interface. Since the IP scanner or high-speed CCD/CMOS are no longer needed, this online FSS system has the advantages of low cost and compactness as well.

We have also shown the functionality of the online FSS prototype through a preliminary experiment. The unfolding spectrum consists of a typical bremsstrahlung spectrum, and the temperature measured is in good agreement with the simulated one in the tens of MeV band. In the future, we will carry out more system test experiments and improve the electromagnetic compatibility, to implement a low-noise system with high unfolding accuracy. With the increasing demands of the high repetition rate and high accuracy, future laser-plasma experiments and application research would benefit from this optimized, integrated, compact, and low-cost online FSS technique.

\section*{Acknowledgements}
The authors thank Zhimeng Zhang and Bo Zhang for their assistance in the experiment. This work is supported by the Natural Science Foundation of China (Grant No. 12004353, 11975214, 11991071, 11905202, 12175212, 12120101005), and Key Laboratory Foundation of The Sciences and Technology on Plasma Physics Laboratory (Grant No.6142A04200103 and 6142A0421010).

\end{document}